# Sub-symmetry protected topological states

Ziteng Wang[1†], Xiangdong Wang[1†], Zhichan Hu[1†], Domenico Bongiovanni[1,2†], Dario Jukić[3], Liqin Tang[1], Daohong Song[1], Roberto Morandotti[2], Zhigang Chen[1*], and Hrvoje Buljan[1,4*]

[1]TEDA Applied Physics Institute and School of Physics, Nankai University, Tianjin 300457, China
[2]INRS-EMT, 1650 Blvd. Lionel-Boulet, Varennes, Quebec J3X 1S2, Canada
[3]Faculty of Civil Engineering, University of Zagreb, A. Kačića Miošića 26, 10000 Zagreb, Croatia
[4]Department of Physics, Faculty of Science, University of Zagreb, Bijenička c. 32, 10000 Zagreb, Croatia
[†]These authors contributed equally to this work
*e-mail: zgchen@nankai.edu.cn, hbuljan@phy.hr

**A hallmark of symmetry-protected topological phases (SPTs) are topologically protected boundary states, which are immune to perturbations that respect the protecting symmetry[1,2]. It is commonly believed that any perturbation that destroys an SPT phase simultaneously destroys the boundary states[1,2]. However, by introducing and exploring a weaker sub-symmetry (SubSy) requirement on perturbations, we find that the nature of boundary state protection is in fact more complex. We demonstrate that the boundary states are protected by only the SubSy using prototypical Su-Schrieffer–Heeger (SSH) and breathing Kagome lattice (BKL) models, even though the overall topological invariant and the SPT phase are destroyed by SubSy preserving perturbations. By employing judiciously controlled symmetry breaking in photonic lattices, we experimentally demonstrate such SubSy protection of topological states. Furthermore, we introduce a long-range hopping symmetry in BKLs, which resolves a debate on the topological nature of their corner states. Our results apply to other systems beyond photonics, heralding the possibility of exploring the intriguing properties of SPT phases in the absence of full symmetry in different physical contexts.**

Symmetry protected topological (SPT) phases of matter are ubiquitous in nature and exist on versatile platforms including condensed matter physics, ultracold atomic gases, and photonics[1,2]. Topological insulators (TIs) induced by spin-orbit coupling, which are protected by time-reversal symmetry (TRS), are a paradigm for SPT phases of matter[1-3]. In topological crystalline insulators, a crystalline point group symmetry protects topological metallic boundary states[1,4].

Imagine an SPT phase with a topological invariant characterizing the bulk states and the associated symmetry protected boundary states. Any perturbation that respects the protecting symmetries will not destroy these boundary states or change the topological invariant without closing the gap between bands[1,2]. However, as pictured in Fig. 1, the situation can be more complex: there are perturbations that preserve the topological invariant but oppose the existence of boundary states[5,6], and vice versa, there are perturbations that leave the boundary states unhurt while destroying the topological invariant[7]. Here we explore the underlying physics for the latter scenario by using the concept of SubSy, where the symmetry equation, involving the Hamiltonian and a symmetry operator, does not hold in the whole Hilbert space, but only holds in its subspace. For a prototypical one-dimensional (1D) SSH model and a BKL model, the SubSys arise from their chiral symmetries, which restrict the possibilities of coupling between sublattices (rigorously defined below). Here, the SubSy means that the symmetry equation holds only on one sublattice. In the case of non-negligible long-range hopping (i.e., non-negligible coupling between distant lattice sites) in BKLs, we find that SubSy and an additional long-range hopping symmetry are sufficient to protect the corner states. Our experiments are performed in photonic structures, which have been established as a fertile platform for exploring novel topological phenomena[8-10]. The main message from our findings is summarized in Fig. 1.

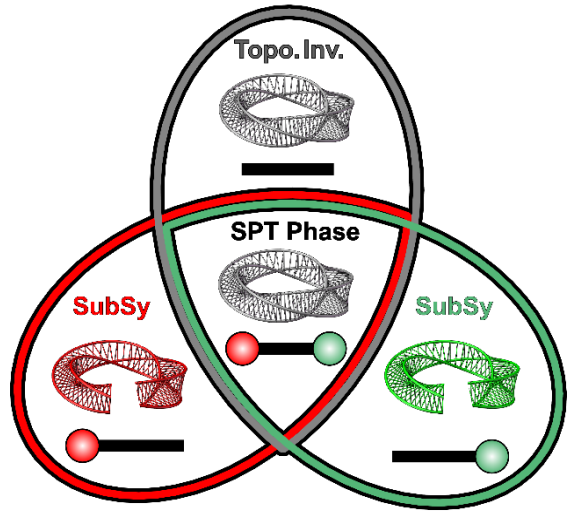

**Fig. 1. Classification of perturbations with respect to symmetries in SPT phases.** A set of perturbations, that preserves the topological invariant (or invariants) and respects a particular symmetry (or symmetries), is encircled with a grey line. Every boundary state is protected with its pertinent sub-symmetry (SubSy). In the illustration, two sets of SubSy-preserving perturbations (encircled with red and green lines) are sketched, but their number depends on the actual system. At the overlap region, one has an SPT phase with a topological invariant characterizing the bulk, and all associated boundary states. For the SSH lattice, the SPT phase is protected by a chiral symmetry (with the implicit assumption of inversion symmetry). The topological invariant is protected by the inversion symmetry (even when the chiral symmetry is broken). The edge state on the A sublattice is protected with A-SubSy, which is defined by a chiral symmetry equation that holds solely on the A sublattice. The same holds for the edge state on the B sublattice with B-SubSy as the protecting subsymmetry. For the BKL with negligible long-range hopping (see text), the $C_3$ symmetry protects the topological invariant, whereas there are three SubSys corresponding to three BKL sublattices, which protect the pertinent higher-order topological corner states.

The SSH lattice illustrated in Fig. 2a represents a typical 1D topological model, originally used to describe polyacetylene[11]. It has subsequently been experimentally realized on versatile platforms including photonics and nanophotonics[12-15], plasmonics[16], quantum optics[17], and in the context of parity-time symmetry and nonlinear non-Hermitian phenomena[18,19].

The SSH lattice is comprised of A and B sublattices (Fig. 2a), with the Hamiltonian $H_{SSH} = \sum_n (t_1 b_n^+ a_n + t_2 a_{n+1}^+ b_n + H.c.)$. Its topological phase is protected by the chiral symmetry,

$$\Sigma_z H_{SSH} \Sigma_z^{-1} = -H_{SSH}, \tag{1}$$

where, $\Sigma_z = P_A - P_B$, and $P_A$ ($P_B$) denotes the projection operator on the A (B) sublattice (see Methods). The system has a trivial phase for $t_1 > t_2$, and a topologically nontrivial phase for $t_1 < t_2$, with the latter being characterized by the Zak phase of $\pi$ and two topologically protected edge states at zero energy (see Fig. 2b). The amplitudes of the left edge state $|A_L\rangle$ are non-zero solely on the A sublattice, that is, $P_A |A_L\rangle = |A_L\rangle$, and $P_B |A_L\rangle = 0$, and analogously for the right edge state $|B_R\rangle$ (see Fig. 2c).

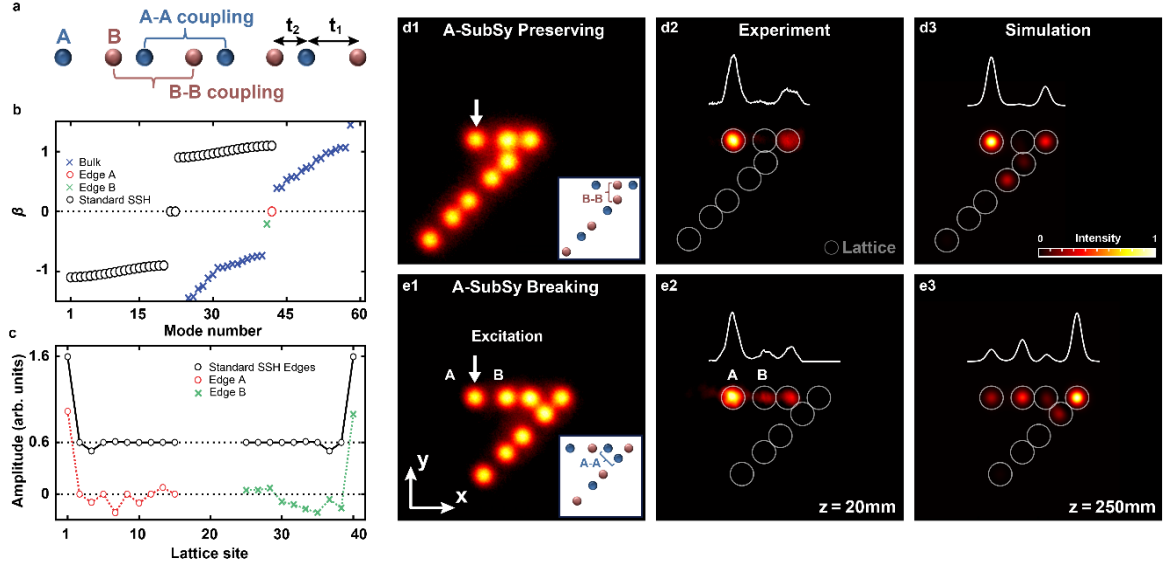

**Fig. 2. Demonstration of the A-SubSy-protected topological state in SSH lattice.** (a) The SSH model. (b) Spectra of the unperturbed (black circles) and the A-SubSy-perturbed system ($t_1 = 0.1, t_2 = 1$). In all subplots, blue crosses are perturbed bands, red circle is the perturbed left edge state at zero energy, and green cross is the perturbed right edge state. (c) Modal structure of the unperturbed (black) and the perturbed left (red) and right (green) edge states. The black modes are shifted up for better visibility. (d1, e1) Experimental images of (d1) A-SubSy-preserving lattice with B-B coupling and (e1) A-SubSy-breaking lattice with A-A coupling. White arrows indicate that solely the left-most waveguide (A sublattice site) is excited at $z = 0$. (d2, e2) Output intensity patterns of the probe beam after propagating 20 mm through corresponding lattices in (d1, e1). The intensity profile in the A-SubSy-preserving lattice resides solely on the A-sublattice (d2), whereas that in the A-SubSy-breaking lattice is present at the B sublattice, which demonstrates SubSy-protection of the left topological edge mode. (d3, e3) Numerical simulations corresponding to (d2, e2), showing output intensity profiles after a much longer propagation distance ($z = 250$ mm).

The concept of SubSy focuses on perturbations that break the chiral symmetry but preserve a less strict SubSy requirement. This provides a theoretical framework and generalizes the partial chiral symmetry breaking case proposed previously in Ref[7]. There are two SubSys in the SSH model, the A-SubSy and the B-SubSy, which are defined by

$$\Sigma_z H_{SSH} \Sigma_z^{-1} P_i = -H_{SSH} P_i, \ i \in \{A, B\}. \tag{2}$$

The most general perturbation of the system implies that the coupling strength between any two lattice sites can be changed (i.e., perturbed) without any restrictions. Any A-SubSy-preserving perturbation can be written as $H' = H_{AB} + H_{BB}$, where $H_{BB}$ denotes the coupling between B sublattice sites (B-B coupling), whereas $H_{AB}$ denotes the coupling between A and B sublattice sites (A-B coupling) (see Methods). Perturbations involving A-A coupling break the A-SubSy.

We consider the influence of A-SubSy-preserving perturbations on the left edge state $|A_L\rangle$. The B-B coupling does not affect this state because $H_{BB}P_A = 0$, which leads to $H_{BB}|A_L\rangle = H_{BB}P_A|A_L\rangle = 0$. However, $H_{BB}$ perturbations break the chiral symmetry, the quantization of the Zak phase, and they affect the right edge state $|B_R\rangle$. Because $H_{AB}$ preserves the chiral symmetry, the edge states are still protected under these perturbations until the gap closes. This is illustrated in Figs. 2b,c which show the spectra and the eigenmode structure for the case of a single randomly chosen A-SubSy-preserving perturbation. The energy of the perturbed left edge mode $|A'_L\rangle$ is intact, but that of the right edge mode as well as the whole spectrum is altered by A-SubSy-preserving perturbations (Fig 2b). The perturbed mode $|A'_L\rangle$ resides solely on the A-sublattice, i.e., $|\langle A'_L|P_A A'_L\rangle|^2 = 1$, however, its structure can differ from the unperturbed mode (see Fig. 2c). Detailed numerical analyses confirm that SubSy requirement is essential for protecting the edge states (see Supplementary Information (SI)).

To experimentally test such edge-state protection with respect to the SubSy-preserving perturbations, we break the chiral symmetry in a controlled fashion. To this end, we introduce the appropriate A-A or B-B hopping by twisting the SSH lattice into the angled structure illustrated in Figs. 2d1 and e1, which either breaks the A-SubSy (Figs. 2e1-e3) or preserves it (Figs. 2d1-d3). The probing is performed by launching a focused beam into the left-most waveguide on the A sublattice. Clearly, the output intensity resides dominantly on the A sublattice (Fig. 2d2), indicating that the left-edge mode is topologically protected when the A-SubSy is preserved. For a direct comparison, in Fig. 2e2 we show the intensity of the same excitation beam propagating through the A-SubSy-breaking lattice. The presence of light in the 2nd waveguide, that is, on the B-sublattice, indicates that it is no longer a topologically protected edge mode,[12,19]. Numerical simulations to even longer propagation distances (Figs. 2d3 and e3) corroborate experimental results.

Kagome lattice is an inexhaustible golden vein of intriguing physics, attracting the broad interest of the scientific community. BKLs, illustrated in Fig. 3a, have been classified as higher-order topological insulators (HOTIs), where topologically protected corner states were observed[20-26]. HOTIs are a new class of topological materials[27], found in condensed matter, networks of resonators, photonic and acoustic systems[20-39]. The corner states in the BKLs were initially considered as HOTI states protected by the generalized chiral symmetry and the $C_3$ crystalline symmetry[20]. However, it was later debated that they are not HOTI states[40-42] because they are not protected by some specific long-range hopping perturbations obeying these symmetries[40]. In our discussion of SubSy-protected corner BKL states we clarify this debated issue.

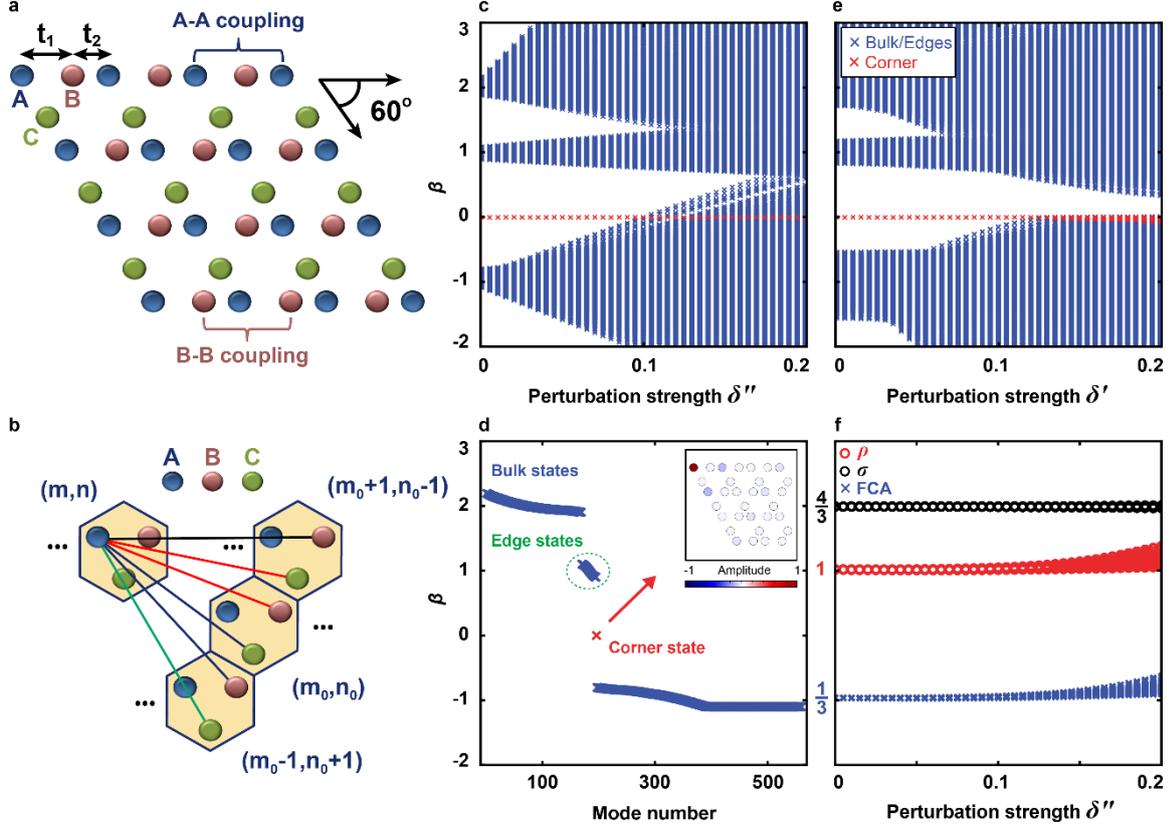

**Fig. 3. Robustness of a Kagome corner state with respect to SubSy perturbations and the LRHS.** (a) Sketch of the rhombic BKL with three sublattices. (b) Illustration of the LRHS condition expressed in Eq. (4), for which the coupling between two sites indicated with red links must be equal, and the same for the coupling indicated with two blue links. (c-e) Eigenvalue spectra of the rhombic BKL flake with 29 lattice sites along one edge for different perturbations. (d) Band gap structure of the unperturbed rhombic BKL flake ($t_1 = 0.1, t_2 = 1$). The rhombic BKL flake has a single corner state shown in the inset. (c) Ensemble of spectra for a set of randomly chosen A-SubSy preserving perturbations $H' = H_{BB} + H_{CC} + H_{BC}$ of various strengths quantified by $\delta'$, which leaves the zero-energy mode (red cross) intact. (e) Corresponding spectra for a set of perturbations $H' + H''$, which respect the A-SubSy and the LRHS ($H'' = H_{AC} + H_{AB}$). The magnitude of perturbations $H'$ is fixed at $\delta' = 0.05$, whereas that of the $H''$ perturbations $\delta''$ is varied. The zero-energy mode (red cross) is protected despite the presence of long-range hopping. In (c) and (e), for each $\delta'$ and $\delta''$ we calculate and plot 70 spectra. (f) Fractional corner anomaly (FCA) and the mode densities ($\rho$ and $\sigma$) calculated for perturbations from (e) – see Methods.

BKLs are comprised of three sublattices (A, B, and C), featuring intracell and intercell hopping amplitudes $t_1$ and $t_2$, respectively (see Fig. 3a and Methods). The bulk polarizations are the topological invariants that characterize the topological phase: for $t_1 < t_2$ the system is in the non-trivial phase with $P_x = P_y = \frac{1}{3}$, whereas for $t_1 > t_2$ the polarizations are zero[20,21]. The BKL Hamiltonian $H_K$ possesses $C_3$ symmetry and the generalized chiral symmetry $\Sigma_3 H_K \Sigma_3^{-1} + \Sigma_3^2 H_K \Sigma_3^{-2} = -H_K$ [20]. Here, $\Sigma_3 = P_A + e^{i\frac{2\pi}{3}} P_B + e^{-i\frac{2\pi}{3}} P_C$, is the symmetry operator, where $P_i, i \in \{A, B, C\}$, are the projection operators. The generalized chiral symmetry yields three equations,

$$\Sigma_3 H_K \Sigma_3^{-1} P_i + \Sigma_3^2 H_K \Sigma_3^{-2} P_i = -H_K P_i, \quad i \in \{A, B, C\}, \tag{3}$$

defining three SubSys corresponding to the three sublattices.

Our theoretical results on BKLs are presented in Fig. 3. We consider a rhombic flake illustrated in Fig. 3a, which has one zero-energy corner state $H_K|A_{cor}\rangle = 0$ residing on the A sublattice, $P_A|A_{cor}\rangle = |A_{cor}\rangle$. The band-gap structure of one such flake is shown in Fig. 3d. First, we consider perturbations between B-B, C-C, and B-C sites, $H' = H_{BB} + H_{CC} + H_{BC}$, which obey the A-SubSy, yet breaking the generalized chiral symmetry. These perturbations obey $H'P_A = 0$, which implies $H'|A_{cor}\rangle = H'P_A|A_{cor}\rangle = 0$, that is, any such perturbation does not affect the corner state. This is illustrated in Fig. 3c which shows the bandgap structure, with the corner state indicated by red crosses, for a set of randomly chosen perturbations $H'$ of various magnitudes quantified by $\delta'$ (see Methods). Interestingly, at some higher perturbation strengths $|A_{cor}\rangle$ can become a bound state in the continuum (BIC)[43].

Next, we consider the A-SubSy-preserving perturbations between A-B and A-C sites: $H'' = H_{AC} + H_{AB}$. Such perturbations can affect the zero-energy corner state $|A_{cor}\rangle$, even in a setup preserving both the full generalized chiral symmetry and the $C_3$ symmetry[40]. Long-range hopping parameters in $H''$ between site A in the unit cell $(m, n)$ and site B in the unit cell $(m_0, n_0)$ are denoted with $t_{ab}^{m,n;m_0,n_0}$, and equivalently for A-C coupling (see Methods). We analytically find that the zero-energy state, residing solely on the A-sublattice, can exist only if the following two conditions hold: $t_{ac}^{m,n;m_0,n_0} = t_{ab}^{m,n;m_0-1,n_0+1}$ and $t_{ac}^{m,n;m_0,n_0} = t_{ab}^{m,n;m_0,n_0}$ (see SI). These conditions are trivially satisfied if the long-range hopping is zero, i.e., when the tight-binding approximation holds. Otherwise, they are too strict and unphysical as illustrated in Fig. 3b, as all coupling strengths between lattice sites indicated with solid lines must be equal.

However, as the coupling strength is typically correlated with the distance, an inspection of the A-B and A-C links in Fig. 3b suggests that, when $t_1 < t_2$, we should explore an approximate, but more physical and less restrictive long-range hopping symmetry (LRHS):

$$t_{ac}^{m,n;m_0,n_0} = t_{ab}^{m,n;m_0-1,n_0+1}. \tag{4}$$

Equation (4) implies that only those couplings indicated by the same color in Fig. 3b must be equal. To test the protection of the corner state under the A-SubSy and LRHS, we calculate the spectra for a set of randomly chosen $H' + H''$ perturbations of different magnitudes quantified by $\delta'$ and $\delta''$, respectively (see Fig. 3e and Methods). We see that the zero-energy corner state remains in the gap and protected, until it is too close to the band at strong perturbations (this is a finite-size effect). The perturbed corner state is dominantly on the A sublattice as long as it is in the gap (see SI).

We experimentally test the protection of the corner state under the SubSy by implementing targeted next-to-nearest neighbor hopping, introduced by imprinting bridge waveguides in the rhombic lattice (see SI for details). As shown in Fig. 4a1, the lattice with B-B and C-C bridges preserves the A-SubSy, while the lattice in Fig. 4b1 with A-A bridges breaks the A-SubSy. For the lattice with a broken A-SubSy, there is light in the B and C waveguides nearest to the corner site (Fig. 4b2). This offers clear evidence that the corner mode is not protected anymore. On the contrary, for the lattice with a preserving A-SubSy, light is present solely on the A sublattice (Fig. 4a2), exhibiting the characteristics of HOTI corner states in BKLs[20,21,39]. This proves that the corner state, in this case, is protected against the B-B and C-C bridge perturbations. To underpin the experimental results, in Figs. 4a3 and b3 we show results from numerical simulations obtained in realistic BKLs with parameters corresponding to those from the experiment, which display an excellent agreement. Long distance simulations also validify that light remains localized as in Fig. 4a3 due to topological protection, but travels through the two bridges in Fig. 4b1 and spread into the bulk (see SI).

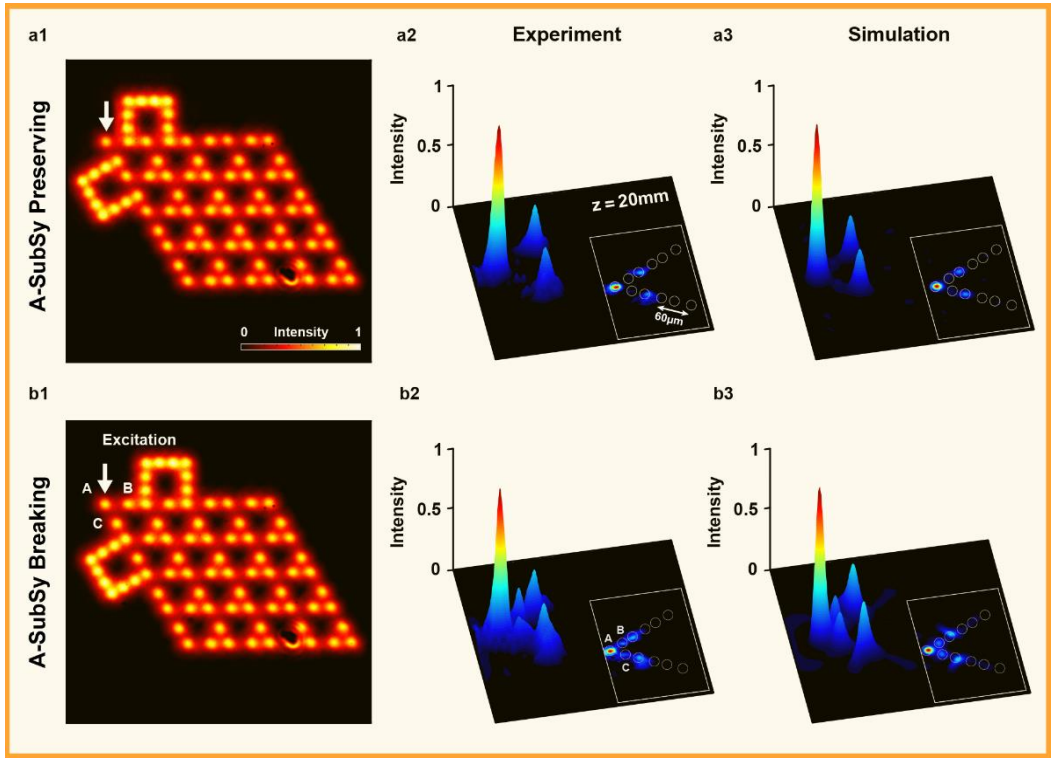

**Fig. 4. Demonstration of the SubSy-protected corner state in rhombic Kagome lattices.** (a1, b1) Experimentally established A-SubSy-preserving lattice (with bridges connecting B-B and C-C sites) and A-SubSy-breaking lattice (with bridges connecting A-A sites). (a2, b2) 3D intensity plots of the output probe beam after propagating through (a2) the A-SubSy-preserving lattice, where light is solely populating the A-sublattice, and (b2) the A-SubSy-breaking lattice, where the presence of light in B and C sublattice sites clearly indicates a destroyed topological corner mode. White arrows in (a1, b1) mark that the corner lattice site initially excited. (a3, b3) Numerically obtained intensity patterns corresponding to the experimental results in (a2, b2), respectively. White circles in the insets of (a2-b3) depict the corner structure of the BKL.

We are now ready to discuss our results with the focus on the diagram in Fig. 1. In the 1D SSH lattice, the left edge mode is protected by A-SubSy (encircled with a red line), while B-

SubSy-preserving perturbations (green) do not affect the right edge mode. At the overlap region, one has the full chiral symmetry and the SPT phase. However, it has been shown that perturbations that respect the inversion symmetry (encircled with a grey line) protect the topological invariant, i.e., the Zak phase, even if the full chiral symmetry is broken[5,6].

The scenario in which BKLs are involved is more complex. First, we consider BKLs where long-range hopping is negligible, which is physically common when hopping is generated with evanescent coupling. The corner state on the A sublattice is robust with respect to A-SubSy-preserving perturbations, and analogously for the corner states on other sublattices (their existence depends on the shape of the BKL flake). The topological invariant is, on the other hand, quantized due to the $C_3$ symmetry[38]. Thus, for a triangular flake of the BKL with $C_3$ and generalized chiral symmetry, one can classify perturbations with respect to symmetries in accordance with Fig. 1 with an additional C-SubSy (not shown), and the grey encircled region corresponding to $C_3$ symmetry. In this model, the BKL corner states are HOTI states.

When the long-range hopping beyond the neighboring unit cells becomes appreciable, the protection of the corner state under A-SubSy and the LRHS can be interpreted as being inherited from the underlying Hamiltonian $H_K$. This interpretation is underpinned by the calculation of the fractional corner anomaly[25] shown in Fig. 3f for an ensemble of random A-SubSy and LRHS-preserving perturbations. It is a clear signature of nontrivial topology and the existence of the corner state.

In conclusion, we have demonstrated SubSy-protected boundary states of SPT phases by employing perturbations that break the original topological invariants. Although the SubSy concept here arises from the chiral symmetries, we envision its applicability for other protecting symmetries as well. For the BKLs with non-negligible long-range hopping, we have unveiled a previously undiscovered LRHS that is essential for protection of the corner states, providing a basis for understanding their HOTI characteristics. Our results extend beyond photonics to condensed matter and cold atom systems, in which many intriguing phenomena are mediated by the interplay of symmetry and topology.

## Methods

**Projection operators.** The projection operator $P_A$ is constructed by requiring that the amplitude of $P_A|\psi\rangle$ is identical to the amplitude of $|\psi\rangle$ on any A-sublattice site, and zero on any other sublattices. The other projection operators ($P_B$, $P_C$) are constructed fully analogously.

**SSH lattice.** The chiral symmetry of the SSH lattice in Eq. (1) implies that for every eigenstate $|e\rangle$ satisfying $H_{SSH}|e\rangle = \beta|e\rangle$, there is another eigenstate $\Sigma_z|e\rangle$ with eigenvalue $-\beta$. This ensures that any perturbation of the Hamiltonian that preserves the chiral symmetry does not destroy the topologically protected edge states unless the gap closes and the system undergoes a topological phase transition to a trivial phase (e.g. see Ref.[2] and refs. therein).

Perturbations of the SSH model corresponding to A-B coupling are formally defined as $H_{AB} = \sum_{m,n}(s_{ab}^{m,n} a_m^\dagger b_n + H.c.)$, where $a_m$ is the annihilation operator at an A sublattice site in the $m$th unit cell, and analogously for $b_n$, while $s_{ab}^{m,n}$ is the strength of the coupling. Similarly, $H_{BB} = \sum_{m,n}(s_{bb}^{m,n} b_m^\dagger b_n + H.c.)$, where $m \neq n$, and analogously for $H_{AA}$.

**Breathing Kagome lattice.** The BKL Hamiltonian is given by

$$H_K = \sum_{m,n}(t_1 a_{m,n}^\dagger b_{m,n} + t_1 a_{m,n}^\dagger c_{m,n} + t_1 b_{m,n}^\dagger c_{m,n} + H.c.) + \sum_{m,n}(t_2 b_{m,n}^\dagger a_{m+1,n} + t_2 c_{m,n}^\dagger a_{m,n+1} + t_2 c_{m,n}^\dagger b_{m-1,n+1} + H.c.),$$

where, $a_{m,n}$ is the annihilation operator at an A sublattice site in the unit cell labeled with $(m,n)$ indices, and analogously for $b_{m,n}$ and $c_{m,n}$. All perturbations between sublattices A and B beyond the hopping described with $t_1$ and $t_2$ can be described by

$$H_{AB} = \sum_{m,n;m_0,n_0}(t_{ab}^{m,n;m_0,n_0} a_{m,n}^\dagger b_{m_0,n_0} + H.c.), (m,n) \neq (m_0,n_0), (m,n) \neq (m_0 - 1, n_0),$$

and analogously for $H_{AC}$ and $H_{BC}$. The B-B hopping perturbations are described by

$$H_{BB} = \sum_{m,n;m_0,n_0}(t_{bb}^{m,n;m_0,n_0} b_{m,n}^\dagger b_{m_0,n_0} + H.c.),$$

and analogously for $H_{AA}$ and $H_{CC}$.

To construct perturbations, the hopping amplitudes between the unit cells $(m,n) \rightarrow (m \pm i, n \pm j)$, $i, j = -3, \ldots, 3$, are perturbed with strength $rand \cdot \delta'|t_2 - t_1|$ for $H' = H_{BC} + H_{BB} + H_{CC}$, and $rand \cdot \delta''|t_2 - t_1|$ for $H'' = H_{AC} + H_{AB}$. The nearest couplings $t_1$ and $t_2$ are not perturbed. All perturbations retain the lattice symmetry. In Fig. 3c, for each magnitude of the perturbation $\delta'$, we calculate and plot an ensemble of 70 spectra for randomly chosen $H'$. An equivalent procedure is used for Fig. 3e, where parameter $\delta''$ is now varied, and $\delta' = 0.05$ is kept fixed. Every $H''$ respects the lattice symmetry, the A-SubSy, and the LRHS. The projections $|\langle A'_{cor}|A_{cor}\rangle|^2$ and $|\langle A'_{cor}|P_A A'_{cor}\rangle|^2$ between the unperturbed $|A_{cor}\rangle$ and the perturbed $|A'_{cor}\rangle$ corner states are exactly unity for any $H' = H_{BB} + H_{CC} + H_{BC}$. However, for $H' + H''$, $|\langle A'_{cor}|A_{cor}\rangle|^2 < 1$ (see SI).

The fractional corner anomaly is calculated as $(\rho - 2*\sigma) \mod 1$ following Ref.[25]; red and black circles in Fig. 3f represent the mode density of the corner unit cell $\rho$ and the average mode density of edge unit cells $\sigma$ (with edges that intersect at the corner), respectively. The mode density is calculated as the local density of states integrated over all states above the bandgap in the propagation constant spectrum (shown in Fig. 3e).

**Experiments.** We establish both 1D SSH and BKL photonic lattices by site-to-site writing of waveguides in a 20-mm-long biased nonlinear crystal with a continuous-wave (CW) laser[19,39]. A low-power ordinarily-polarized laser beam at 532 nm illuminates a spatial light modulator (SLM), which creates a lattice-writing beam with variable spatial input positions. All waveguides remain intact during measurements. Probing of topological edge/corner states is performed with a low power (20nW) extraordinarily-polarized beam, which undergoes linear propagation through the lattices (see SI for details).

**Acknowledgments**

We acknowledge assistance from Ruoqi Cheng and Yihan Wang. This research is supported by the National Key R&D Program of China under Grant No. 2017YFA0303800, the National Natural Science Foundation (12134006, 11922408) and the QuantiXLie Center of Excellence, a project co-financed by the Croatian Government and European Union through the European Regional Development Fund - the Competitiveness and Cohesion Operational Programme (Grant KK.01.1.1.01.0004). D. B. acknowledges support from the 66 Postdoctoral Science Grant of China. R.M acknowledge support from NSERC and the CRC program in Canada

# Supplementary Information for

# Sub-symmetry protected topological states

Ziteng Wang[1†], Xiangdong Wang[1†], Zhichan Hu[1†], Domenico Bongiovanni[1,2†], Dario Jukić[3], Liqin Tang[1], Daohong Song[1], Roberto Morandotti[2], Zhigang Chen[1*], and Hrvoje Buljan[1,4*]

[1]TEDA Applied Physics Institute and School of Physics, Nankai University, Tianjin 300457, China
[2]INRS-EMT, 1650 Blvd. Lionel-Boulet, Varennes, Quebec J3X 1S2, Canada
[3]Faculty of Civil Engineering, University of Zagreb, A. Kačića Miošića 26, 10000 Zagreb, Croatia
[4]Department of Physics, Faculty of Science, University of Zagreb, Bijenička c. 32, 10000 Zagreb, Croatia
[†]These authors contributed equally to this work
*email: zgchen@nankai.edu.cn, hbuljan@phy.hr

## 1. Numerical demonstration of SubSy-protected topological edge states: SSH lattices

Let us supplement our results on the SubSy-protected edge states in the SSH model presented in Fig. 2 of the main text with numerical simulations. To test the protection of the left edge state $|A_L\rangle$ with respect to A-SubSy-preserving perturbations, we perturb $H_{SSH}$ with $H_{BB}$ (B-B coupling) and $H' = H_{AB} + H_{BB}$ (A-B and B-B coupling). In our numerical simulations, the coupling amplitudes for $i \to i \pm n$, where $n = 2, \ldots, 6$, are perturbed with strength $rand \cdot \delta |t_2 - t_1|$, where $rand \in [0,1]$ is chosen at random, and a dimensionless parameter $\delta$ quantifies the strength of the perturbation. All perturbations maintain the lattice periodicity.

In Fig. S1a, for each value of $\delta$, we plot superimposed spectra for a set of 70 random $H_{BB}$ perturbations. The scalar products $|\langle A'_L | A_L \rangle|^2$ and $|\langle A'_L | P_A A'_L \rangle|^2$ (detailed below) are exactly unity for all these perturbations, which means that $H_{BB}$ perturbations do not affect the left edge state at all. In Fig. S1b we show spectra for a set of 70 randomly chosen $H' = H_{AB} + H_{BB}$ perturbations. In this case, the energy of the left edge mode (red circles) still remains to be zero in all plots, although for these $H'$ perturbations the structure of the left edge mode changes, which is quantified by $|\langle A'_L | A_L \rangle|^2 < 1$ (see Fig. S2 below). In both cases, whether with only B-B coupling perturbations or with joint A-B and B-B coupling perturbations, the right edge state $|B_R\rangle$ (blue crosses in Fig. S1) is strongly affected, driven away from the zero-energy position.

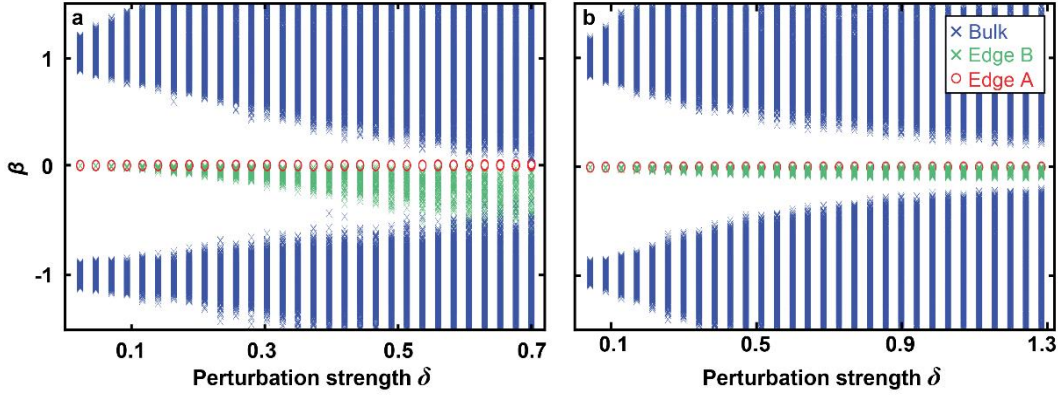

**Fig. S1. Numerical demonstration of SubSy-protected topological edge states in the SSH lattice.** In both plots, blue crosses indicate perturbed bands, red circles denote the perturbed left edge state at zero energy, and green crosses denote the perturbed right edge state. (a) Superimposed spectra for random B-B coupling perturbations (70 spectra for each $\delta$). (b) The same as (a) but for joint A-B and B-B (i.e., general A-SubSy) coupling perturbations. The eigenvalue of the left edge state is always at zero energy, for any random A-SubSy-preserving perturbation, in contrast with that of the right edge state. These results validate the topological protection under the SubSy.

**Scalar products and eigenstates of the projection operators:**

Next, we provide additional information on the scalar products $|\langle A'_L|A_L\rangle|^2$ and $|\langle A'_L|P_A A'_L\rangle|^2$ for the numerical simulations corresponding to Fig. S1b. The scalar product $|\langle A'_L|A_L\rangle|^2$ between the unperturbed $|A_L\rangle$ and the perturbed $|A'_L\rangle$ left edge states is shown with red circles in Fig. S2; it quantifies the change in the mode profile of the edge state under perturbation. The scalar product $|\langle A'_L|P_A A'_L\rangle|^2$, shown with blue crosses in Fig. S2 quantifies the amount of the modal amplitude that resides on the A sublattice; if $|\langle A'_L|P_A A'_L\rangle|^2 = 1$, then 100% of the amplitude is on the A sublattice. Both scalar products are plotted as a function of the perturbation strength $\delta$. We see that the mode profile of the perturbed edge state can deviate from the unperturbed one, but it nevertheless possesses zero eigenvalue (see Fig. S1b) and it always remains localized at the left edge (see Fig. 2c of the main text). From $|\langle A'_L|P_A A'_L\rangle|^2$ we clearly see that the perturbed state resides solely on the A-sublattice until perturbations are sufficiently strong so that the finite-size effects kick in (but even in this latter case this deviation is very small, i.e., $|\langle A'_L|P_A A'_L\rangle|^2 \approx 1$).

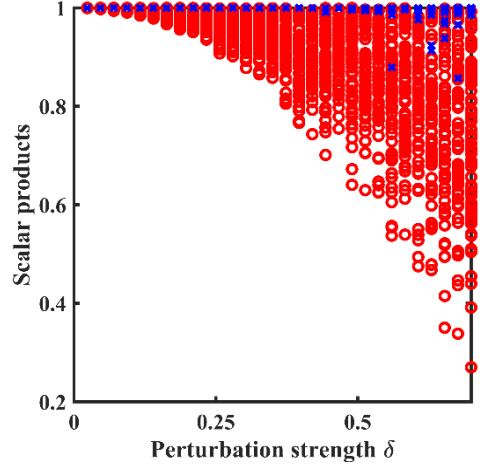

**Fig. S2. Scalar products for A-SubSy-preserving perturbations of the SSH lattice.** Scalar products $|\langle A'_L|A_L\rangle|^2$ (red circles) and $|\langle A'_L|P_A A'_L\rangle|^2$ (blue crosses), as functions of the perturbation strength $\delta$, for the numerical simulations corresponding to Fig. S1b.

## 2. Long-range hopping symmetry (LRHS) and scalar product calculations: BKLs

In this section, we first provide an analytical derivation of the conditions for the LRHS. Let us consider the breathing Kagome lattice (BKL) Hamiltonian (see Refs.[1-7] and references therein)

$$H_K = \sum_{m,n}(t_1 a^\dagger_{m,n} b_{m,n} + t_1 a^\dagger_{m,n} c_{m,n} + t_1 b^\dagger_{m,n} c_{m,n} + H.c.) + \sum_{m,n}(t_2 b^\dagger_{m,n} a_{m+1,n} +$$
$$+ t_2 c^\dagger_{m,n} a_{m,n+1} + t_2 c^\dagger_{m,n} b_{m-1,n+1} + H.c.), \qquad (S1)$$

where, $a_{m,n}$ is the annihilation operator at the A site in the unit cell labeled with $(m,n)$ indices, and analogously for $b_{m,n}$ and $c_{m,n}$. Long-range hopping amplitudes are denoted with $t_{ab}^{m,n;m_0,n_0}$ for the coupling between site A in the unit cell $(m,n)$ and site B in the unit cell $(m_0, n_0)$, and analogously for the B-C and A-C coupling (see the main text and Methods). All perturbations between sublattices A and B (not including those denoted with $t_1$ and $t_2$) can be described by $H_{AB} = \sum_{m,n;m_0,n_0}(t_{ab}^{m,n;m_0,n_0} a^\dagger_{m,n} b_{m_0,n_0} + H.c.)$, $(m,n) \neq (m_0, n_0)$, $(m,n) \neq (m_0 - 1, n_0)$, and analogously for $H_{AC}$ and $H_{BC}$. The B-B hopping perturbations are described by $H_{BB} = \sum_{m,n;m_0,n_0}(t_{bb}^{m,n;m_0,n_0} b^\dagger_{m,n} b_{m_0,n_0} + H.c.)$, and analogously for $H_{AA}$ and $H_{CC}$. Following the main text, we denote perturbations with $H' = H_{BB} + H_{CC} + H_{BC}$ and $H'' = H_{AC} + H_{AB}$. Perturbation $H'$ does not affect the A sublattice subspace but perturbation $H''$ does; however, both cases do not break the A-SubSy.

Therefore, we seek for an eigenstate of $H_K + H''$ with zero eigenvalue that resides solely on the A sublattice: $|\psi_A\rangle = \sum_{m,n} A_{m,n} a^\dagger_{m,n}|0\rangle$. From $(H_K + H'')|\psi_A\rangle = 0$ we obtain the following equations

$$t_1 A_{m_0,n_0} + t_2 A_{m_0+1,n_0} + \sum_{m,n} t_{ab}^{m,n;m_0,n_0} A_{m,n} = 0.$$
$$t_1 A_{m_0,n_0} + t_2 A_{m_0,n_0+1} + \sum_{m,n} t_{ac}^{m,n;m_0,n_0} A_{m,n} = 0. \quad (S2)$$

Let us now replace $n_0$ with $n_0 + 1$ in the first equation in (S2), and $m_0$ with $m_0 + 1$ in the second equation in (S2) to obtain

$$t_1 A_{m_0,n_0+1} + t_2 A_{m_0+1,n_0+1} + \sum_{m,n} t_{ab}^{m,n;m_0,n_0+1} A_{m,n} = 0,$$
$$t_1 A_{m_0+1,n_0} + t_2 A_{m_0+1,n_0+1} + \sum_{m,n} t_{ac}^{m,n;m_0+1,n_0} A_{m,n} = 0, \quad (S3)$$

Next, by subtracting the two equations in (S2) and (S3), we get:

$$t_2(A_{m_0,n_0+1} - A_{m_0+1,n_0}) + \sum_{m,n} t_{ac}^{m,n;m_0,n_0} A_{m,n} - \sum_{m,n} t_{ab}^{m,n;m_0,n_0} A_{m,n} = 0,$$
$$t_1(A_{m_0,n_0+1} - A_{m_0+1,n_0}) + \sum_{m,n} t_{ab}^{m,n;m_0,n_0+1} A_{m,n} - \sum_{m,n} t_{ac}^{m,n;m_0+1,n_0} A_{m,n} = 0. \quad (S4)$$

We combine the two equations in (S4) and we get

$$t_1\left(\sum_{m,n} t_{ac}^{m,n;m_0,n_0} A_{m,n} - \sum_{m,n} t_{ab}^{m,n;m_0,n_0} A_{m,n}\right) - t_2\left(\sum_{m,n} t_{ab}^{m,n;m_0,n_0+1} A_{m,n} - \sum_{m,n} t_{ac}^{m,n;m_0+1,n_0} A_{m,n}\right) = 0 \quad (S5)$$

Equation (S5) holds for every $(m_0, n_0)$. For the sake of clarity, let us write down Eq. (S5) for $(m,n) = (1,1)$ and $m_0, n_0 \leq 2$:

$$-t_2(t_{ab}^{1,1;1,2} - t_{ac}^{1,1;2,1}) = 0,$$
$$t_1(t_{ac}^{1,1;2,1} - t_{ab}^{1,1;2,1}) - t_2(t_{ab}^{1,1;2,2} - t_{ac}^{1,1;3,1}) = 0,$$
$$t_1(t_{ac}^{1,1;1,2} - t_{ab}^{1,1;1,2}) - t_2(t_{ab}^{1,1;1,3} - t_{ac}^{1,1;2,2}) = 0. \quad (S6)$$

By exploring Eqs. (S5) and (S6), we arrive at the mathematical condition that the long-range hopping needs to obey for $|\psi\rangle = \sum_{m,n} A_{m,n} a_{m,n}^\dagger |0\rangle$ to be the zero-energy eigenstate:

$$t_{ac}^{m,n;m_0,n_0} = t_{ab}^{m,n;m_0,n_0},$$
$$t_{ab}^{m,n;m_0,n_0+1} = t_{ac}^{m,n;m_0+1,n_0}. \quad (S7)$$

The 2$^{nd}$ equation in (S7) is readily expressed as $t_{ab}^{m,n;m_0-1,n_0+1} = t_{ac}^{m,n;m_0,n_0}$.

By exploring the two conditions in Eq. (S7), we find that while mathematically they are perfectly sensible, they are physically reasonable only when the long-range hopping is negligible, i.e., when $t_{ac}^{m,n;m_0,n_0} = t_{ab}^{m,n;m_0,n_0} = t_{ab}^{m,n;m_0-1,n_0+1} = 0$ (see the discussion and illustration in Fig. 3b of the main text).

The zero-eigenvalue eigenstate $|\psi_A\rangle$ of $H_K$ exists only in the topologically nontrivial regime with $t_1 < t_2$. A geometrical inspection of the A-B and A-C links in Fig. 3b of the main text (see red and blue colored links) suggests that, when $t_1 < t_2$, the relation $t_{ac}^{m,n;m_0,n_0} = t_{ab}^{m,n;m_0-1,n_0+1}$ is in fact a more physical condition because the pertinent B and C sites are closer. For this reason, we define the long-range hopping symmetry (LRHS) with Eq. (4) in the main text.

**Scalar products for BKLs:**

In this subsection, we provide additional information on the scalar products $|\langle A'_{cor}|A_{cor}\rangle|^2$ and $|\langle A'_{cor}|P_A A'_{cor}\rangle|^2$ for the numerical simulations corresponding to Fig. 3c and Fig. 3e of the main text, for the breathing Kagome lattice. First, we analyze the results from Fig. 3c, which show superimposed spectra for an ensemble of 70 randomly chosen $H' = H_{BC} + H_{BB} + H_{CC}$ A-SubSy-preserving perturbations quantified by magnitude $\delta'$. As seen from our calculations presented in Fig. S3a, the scalar products are exactly unity, clearly showing that these A-SubSy-preserving perturbations of this type do not affect the corner state at all.

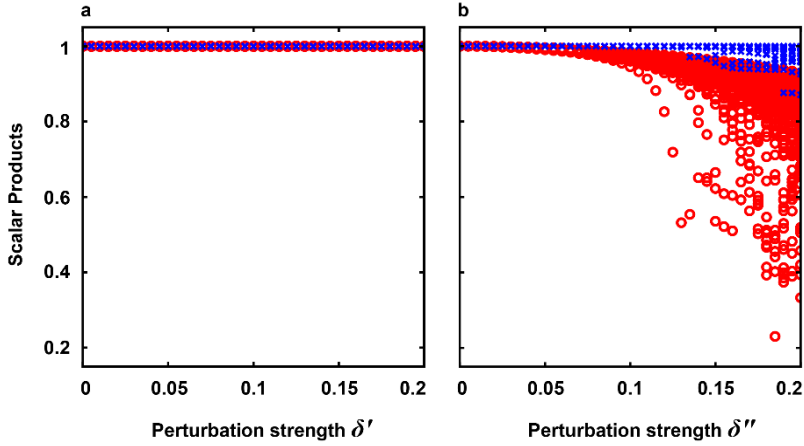

**Fig. S3. Scalar products for A-SubSy-preserving perturbations of the BKL.** (a) Scalar products $|\langle A'_{cor}|A_{cor}\rangle|^2$ (red circles) and $|\langle A'_{cor}|P_A A'_{cor}\rangle|^2$ (blue crosses), as functions of the perturbation strength $\delta'$ for the numerical simulations corresponding to Fig. 3c of the main text, showing that these A-SubSy-preserving perturbations do not affect the mode profile of the corner state and its distribution in the A sublattice. (b) Scalar products as functions of the perturbation strength $\delta''$ for the numerical simulations corresponding to Fig. 3e of the main text. In this latter case, at a sufficiently large perturbation, the finite size-effects become important, and the profile of the perturbed corner state deviates significantly from the unperturbed one; however, it still remains localized dominantly on the A sublattice.

Next, we present the scalar products for Fig. 3e of the main text, which shows superimposed spectra for an ensemble of 70 randomly chosen $H' + H''$ perturbations. The magnitude of the $H'$ component was held fixed at $\delta' = 0.05$, whereas the magnitude of the $H'' = H_{AC} + H_{AB}$ component was varied as quantified by the parameter $\delta''$ (see Fig. 3e of the main text). The calculated quantities $|\langle A'_{cor}|A_{cor}\rangle|^2$ and $|\langle A'_{cor}|P_A A'_{cor}\rangle|^2$ for these perturbations are shown in Fig. S3b. We see that, in this case, the perturbed corner state deviates from the unperturbed corner state as the magnitude of the perturbation is increased. However, it remains localized dominantly on the A sublattice and at the corner until its zero eigenvalue becomes adjacent to the band (see Fig. 3e in the main text), which is a finite size effect (i.e., it happens because the BKL rhombic flake is of finite size).

## 3. Experimental setup and methods

In our experiments, we establish the desired photonic lattices (either the 1D "angled" SSH lattice as shown in Fig. 2 or the 2D rhombic Kagome lattice as shown in Fig. 4 of the main text) by site-to-site writing of waveguides in a strontium-barium niobate (SBN:61) photorefractive (PR) crystal with a continuous-wave (CW) laser[8,9]. As illustrated in Fig. S4, a low-power laser beam at 532nm wavelength illuminates a spatial light modulator (SLM), which creates a quasi-non-diffracting writing beam with variable input positions to the 20mm-long biased crystal. The lattice-writing beam is ordinarily-polarized, while the probe beam launched to the lattice edge is extraordinarily-polarized. Because of the self-focusing nonlinearity and the PR "memory" effect, all waveguides remain intact during the writing and subsequently probing processes. Compared to the femtosecond laser-writing method largely employed in glass materials[7], the photonic lattices in our crystal can be readily reconfigured from topological nontrivial to trivial structures simply by controlling the lattice spacing. After the multi-step writing process (with a bias field of 130kV/m) is completed, the whole lattice can be examined by sending a set of Gaussian beams into the crystal to probe the waveguides one by one, which leads to superimposed lattice structures as shown in Fig. 2 and Fig. 4. In order to investigate the evolution of the topological states in this work, the probing beam used to excite the lattice edge/corner is set at a much weaker power of only about 20nW, so it undergoes only linear propagation without nonlinear self-action through the lattice. (Of course, if needed, the probe power can be increased to locally change the index structure of the lattices – the ingredient used for nonlinear control of topological states as in our previous work[8,9]). The intensity patterns of the probe beam exiting the lattices [Fig. 2 and Fig. 4] are captured by an imaging lens together with a CCD camera.

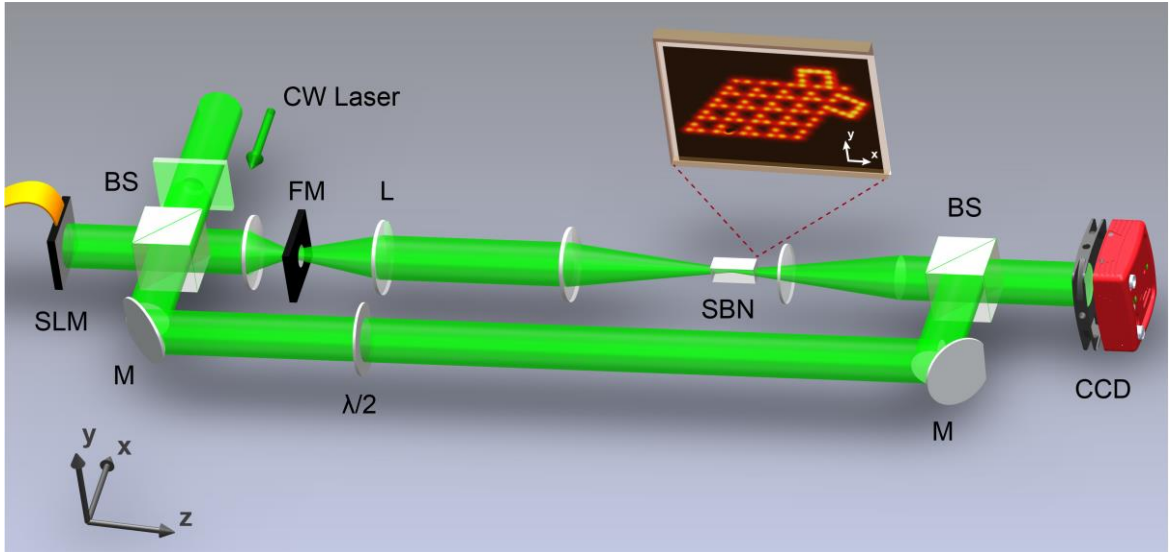

**Fig. S4. Schematic illustration of the experimental setup employed for writing and probing a photonic lattice in a photorefractive crystal.** SLM: spatial light modulator; BS: beam splitter; FM: Fourier mask; L: circular lens; SBN: strontium barium niobite crystal; M: mirror; $\lambda/2$: half-wavelength plate; CCD: charge-coupled device.

## 4. Numerical beam propagation simulations using the continuous model

The propagation of light in a photonic lattice as experimentally implemented in the PR crystal can be described by the following nonlinear Schrödinger-like equation (NLSE) (e.g., see Ref.[9] and references therein):

$$i\frac{\partial \psi}{\partial z} + \frac{1}{2k}\left(\frac{\partial^2 \psi}{\partial x^2} + \frac{\partial^2 \psi}{\partial y^2}\right) + \frac{k\Delta n}{n_0}\frac{\psi}{1+I_{Lattice}(x,y)+I_{Probe}(x,y,z)} = 0, \quad (S8)$$

where $\psi(x, y, z)$ is the electric field envelope, with $x$ and $y$ being the transverse coordinates, $z$ the propagation distance, and $k = 2\pi n_0/\lambda_0$ is the wavenumber in the PR crystal, with $\lambda_0$ denoting the laser wavelength in the vacuum. For the specific SBN crystal we used, the refractive index change is modeled as $\Delta n = -n_0^3 r_{33} E_0/2$, where $n_0 = 2.35$ is the crystal index, $r_{33} = 280$ pm/V is the electro-optic coefficient along the crystalline $c$-axis, and $E_0$ is the biased electric field. $I_{Lattice}(x,y)$ is the intensity of the lattice-writing beam, and $I_{Probe}(x,y,z)$ is the intensity of the probe beam, which is very weak to guarantee a linear propagation once the lattice is written, i.e., $I_{Lattice}(x,y) + I_{Probe}(x,y,z) \approx I_{Lattice}(x,y)$. By using Eq. (S8), we perform numerical simulations for corner excitations in both the 1D SSH lattice and the 2D BKL using parameters close to those from the experiments.

Typical simulation results for the BKLs are shown in Fig. S5, comparing the corner excitation for three different cases, especially the "bridge" effect on preserving or breaking the A-SubSy. As shown in Figs. S5a2 and a3, the probe beam evolves into a characteristic topological corner mode in the standard BKL without the bridge, populating mainly the corner A site and its next-nearest-neighbor A sublattice sites (but not the B and C sublattice sites)[1,2]. When two bridges of waveguides are added to introduce long-range coupling in A sublattice as shown in Fig. S5c1, a large amount of light travels through the bridges as well as along the lattice edges (see Figs. S5c2-c3) and also media files in SM), thus breaking the A-SubSy. As a result, the topological corner mode cannot persist, with an evident coupling of light into the B and C sublattice sites.

On the other hand, when the two bridges introduce B-B and C-C coupling (see Fig. S5b1), they still preserve the A-SubSy. In this case, the probe beam evolves into a topological corner mode, with no light in the B and C sites closest to the corner A site (Figs. S5b2-b3), in sharp contrast to the A-SubSy-breaking case (Figs. S5c2-c3). Remarkably, our numerical results show that light propagation in the A-SubSy-preserving BKL with such bridges resembles that in the standard BKL. In all cases, numerical simulations are performed by probing only the corner waveguide and then letting the probe beam evolve during propagation. These results clearly prove the feasibility of using waveguide bridges to introduce long-range coupling for controlled SubSy-preserving or -breaking in the photonic lattice, even in frameworks where long-range coupling cannot be easily achieved otherwise.

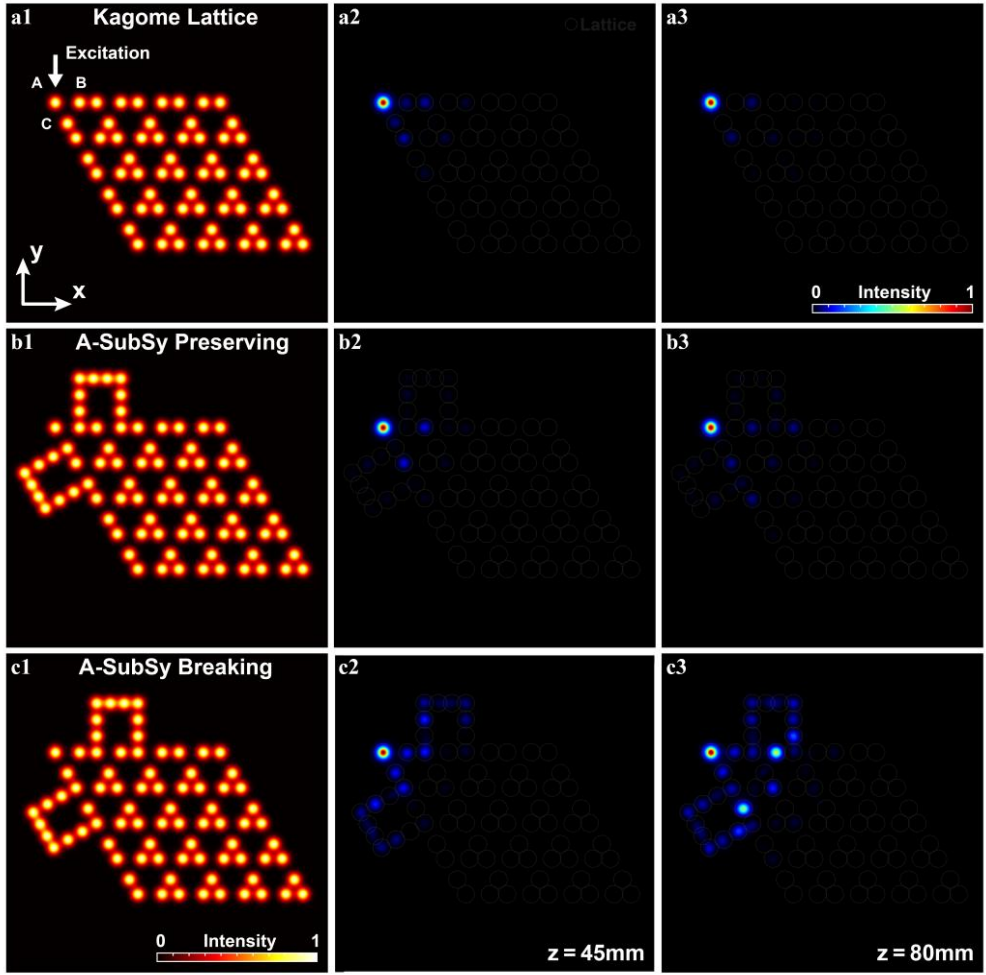

**Fig. S5. Numerical simulation of corner excitation in A-SubSy-preserving/breaking BKLs.** We show light propagation under single-site corner excitation in (a1-a3) standard nontrivial BKL (no bridge), (b1-b3) A-SubSy-preserving BKL (with bridges connecting B sites and C sites), and (c1-c3) A-SubSy-breaking BKL (with bridges connecting A sites). (a1-c1) Photonic lattice structures; (a2-c2) and (a3-c3) Output intensity patterns of the probe beam retrieved at $z = 45$ mm and 80 mm, respectively. Notice the dramatic difference at the output when the bridge is shifted (see supplementary media files for the evolution of the beam dynamics): It is evident that light does not travel through the bridges in (b1) due to topological protection of the corner state, but it takes the longer paths through the two bridges in (c1) although the bridges are now a bit further away from the corner compared to those in (b1).